\documentclass{article}[12pt]
\usepackage[pdftex]{graphicx}   
\begin{document}
\title{Extracting vacuum expectation values from approximate vacuum prepared by the adiabatic quantum computation}
\author{Kazuto Oshima\thanks{E-mail: kooshima@gunma-ct.ac.jp}    \\ \\
\sl National Institute of Technology, Gunma College,Maebashi 371-8530, Japan}
\date{}
\maketitle
\begin{abstract}
We propose a procedure to extract vacuum expectation values from approximate vacuum prepared by the adiabatic quantum computation.
We use plural ancilla bits with hierarchical structure, intending to gradually put up approximate precision.
We exhibit simulation results for a typical one-qubit system and a two-qubits system based on the (1+1)-dimensional Schwinger model using
classically emulated digital quantum simulator. 
\end{abstract} 
\\
\newpage
\section{Introduction}
In quantum mechanics the quantum adiabatic theorem\cite{Born} is well known.    According to this theorem, if an initial state is a ground state of a
time-changing Hamiltonian, the state remains to be a ground state of the Hamiltonian, as far as the Hamiltonian changes sufficiently gradually in time.
The quantum adiabatic theorem has been used by Nishimori and Kadowaki \cite{Nishimori} to find ground states of Ising models.   They have
adopted a transverse Ising model as the initial Hamiltonian, whose ground state is trivial.    Starting from a trivial ground state of a simple
Hamiltonian, we can obtain a non-trivial ground state of a complicated Hamiltonian.   This procedure is called as the adiabatic quantum 
computation.   Farhi et. al. \cite{Farhi1} have advocated to use the adiabatic quantum computation to solve combination optimization problems
such as 3-SAT. 

Since the publication of the universal type of quantum computer, some quantum systems have been analyzed by the universal type
of quantum computer\cite{Martinez1}.     Recently, the quantum simulator also has been used to analyze quantum systems.    The quantum simulator is
free from noises and is suited for precise alalysis of quantum systems.   The (1+1)-dimensional Schwinger model\cite{Schwinger} has been
analyzed by the quantum simulator\cite{Honda1,Honda2, Okuda}.   By virtue of the analysis of the quantum simulator, an unknown region of parameters for the  (1+1)-dimensional 
Schwinger mode has been explored.   If  established results can be reproduced the analysis by the quantum simulator will be reliable.
Some exact results are known for the (1+1)-dimensional Schwinger model in the massless case\cite{Coleman}.    For the (1+1)-dimensional
Schwinger model, in the massless case the vacuum expectation value of ${\bar \psi}\psi$ has been computed as  $\langle {\rm vac} |{\bar \psi} \psi|{\rm vac} \rangle=-{e^{\gamma} \over 2\pi}{g \over \sqrt{\pi}} \approx -0.160g$\cite{Adam},  
where $\gamma \approx 0.5772$ is the Euler constant and  $g$ is the coupling constant between the fermionic fields and the electric field.
Chakraborty et. al.\cite{Honda1} have computed the value $\langle {\rm vac} |{\bar \psi} \psi|{\rm vac}\rangle$ by an extrapolation with respect to the number of qubits.
Their result is consistent with the exact result within a kind of error.

It also has been recognized that an approximate vacuum prepared by the adiabatic quantum computation slightly differs from a true vacuum\cite{Honda1}.
Therefore, vacuum expectation values of some physical quantities by this approximate vacuum are affected  by exited states and slightly differ 
from exact values and their variances are not small.   Recently, the author of the present paper has proposed a procedure to improve the approximate vacuum\cite{Oshima}.  
  The influence of excited states has been diminished and the variances of the vacuum expectation values  have been suppressed.
In the previous paper, we have used  a priori an exact value of ground state energy to improve the approximate vacuum. 
In this point, the previous procedure is not self contained.
 
In this paper, we propose a procedure to estimate vacuum expectation values of physical quantities precisely from the approximate
vacuum prepared by the adiabatic quantum computation.    We use multiple ancilla qubits to exclude iteratively excited states from 
the approximate vacuum.  This time, we do not a priori use the exact value of the ground state energy.  We use an approximate
value of a ground state energy estimated from the approximate vacuum to improve the approximate vacuum.   We update
the approximate value of a ground state energy using the improved approximate vacuum.   We can iterate this improving procedure
as the number of ancilla qubits.    We estimate vacuum expectation values of physical quantities from the iteratively improved 
approximate vacuum.   We examine a one-qubit system and a two-qubits system that leads to (1+1)-dimensional Schwinger model.
\\ \\
\section{One-qubit system}
\subsection{ Hamiltonian and approximate vacuum prepared by the adiabatic quantum computation}
We consider the following simple one-qubit Hamiltonian
\begin{equation}
{\hat H}_{T}=X+JZ,
\end{equation}
where $J$ is a real parameter.   We mainly concentrate on the typical case $J=1$, which will be the easiest case to theoretically analyze.   To prepare a vacuum of ${\hat H}$ by the adiabatic quantum computation
we introduce the initial Hamiltonian ${\hat H}_{0}=Z$.    We have the following Hamiltonian ${\hat H}_{A}(s)$ to carry out the quantum adiabatic computation
\begin{equation}
{\hat H}_{A}(s)=(1-s){\hat H}_{0}+s{\hat H}_{T}=(1-s)Z+s(X+JZ), 
\end{equation}
where $s$ is a parameter that varies gradually from $s=0$ to $s=1$.   We set $s={t \over T}$ for an adequate time period $T$.  We start from the initial state $|1\rangle$ that is the ground state of ${\hat H}_{0}$.
We compute a vacuum expectation value of $Z$ at each $s$.    
We show a simulation result in Fig.1.     After the time $t=T$, the system time develops under the constant Hamiltonian $H_{T}$.     If the true ground state is prepared the
vacuum expectation value of $Z$ should be a constant.      In the result in Fig.1 the vacuum expectation value of $Z$ oscillates regularly.  This means that the prepared
state by the present adiabatic quantum  computation is not a true ground state.    In computer simulation we cannot use a continuous time,
which will cause deviation from a true vacuum.   In the following we examine the approximate vacuum prepared by the adiabatic quantum computation for the case 
$J=1$. 

For $J=1$, the ground state of
$H_{T}$ is 
\begin{equation}
|E_{0}\rangle={1 \over \sqrt{4+2\sqrt{2}}}(|0\rangle-(\sqrt{2}+1)|1\rangle),
\end{equation}
and the first excited state of $H_{T} $ is
\begin{equation}
|E_{1} \rangle={1 \over \sqrt{4-2\sqrt{2}}}(|0\rangle+(\sqrt{2}-1)|1\rangle).
\end{equation}
\\
We see that $\langle E_{0}|H_{T}|E_{0}\rangle=-\sqrt{2}$ and $\langle E_{0}|Z|E_{0}\rangle=-{\sqrt{2} \over 2}$.

After the adiabatic state preparation process, the observable $Z$ time develops as 
\begin{equation}
e^{i{\hat H}_{T}t}Ze^{-i{\hat H}_{T}t}=e^{-iJ{H}t}Ze^{iJ{H}t}={1 \over {2}}(X+Z)+{1 \over \sqrt{2}}Y\sin{2\sqrt{2}t}+{1 \over 2}(Z-X)\cos{2\sqrt{2}t}. 
\end{equation}
At the time $t=T$ if we have a state $|\psi(t=T)\rangle=|\psi(0)\rangle=|\psi_{0}\rangle=\alpha|E_{0}\rangle+\beta|E_{1}\rangle, |\alpha|^{2}+|\beta|^{2}=1$, instead of the 
desired state $|E_{0}\rangle$, we have at a time $t(\ge T)$
\begin{equation}
\langle \psi(t)|Z|\psi(t)\rangle=-{1 \over \sqrt{2}}(1-2|\beta|^{2})+\sqrt{2}|\alpha\beta|\cos(2\sqrt{2}(t-T)+\theta),
\end{equation}
where we have again set $\alpha\beta^{*}= |\alpha\beta^{*}|e^{i\theta}$.   A procedure to obtain an approximate vacuum expectation value of $Z$
is to take time average of $\langle \psi(t)|Z|\psi(t)\rangle$ over the time region $t \ge T$.   In this procedure, however, the approximate value
is always accompanied with the $O(|\beta|^{2})$ systematic error, although it may be small compared with  $O(|\beta|)$.    We propose 
another attempt.

To extract vacuum expectation values of physical quantities from the approximate vacuum $|\psi_{0}\rangle$, we introduce multiple ancilla qubits in the
state $|0\rangle|0\rangle\cdots|0\rangle$.
Our purpose is transform the state $(\alpha|E_{0}\rangle+\beta|E_{1}\rangle)|0\rangle$ into the ideal state $\alpha|E_{0}\rangle|0\rangle+\beta|E_{1}\rangle|1\rangle$.
We compute an approximate vacuum energy $E_{00}$ from the approximate vacuum  $|\psi_{0}\rangle$ by  $E_{00}=\langle \psi_{0}|{\hat H}_{T}|\psi_{0}\rangle$.
We compute a time parameter $\theta_{0}$ that satisfies $\theta_{0}E_{00}={\pi \over 2}$.    We transform the first ancilla bit by the Hadamard operator $H$.
Taking the first ancilla bit as the control bit, we operate $U(\theta_{0})=ie^{-i\theta_{0}{\hat H}_{T}}$ on the physical state $|\psi_{0}\rangle$.
We Hadamard transform the first ancilla bit.  We call this series of operations as a twirling operation.  By this twirling operation, we can partially exclude the excited state.  The initial state
 $(\alpha|E_{0}\rangle+\beta|E_{1}\rangle)|0\rangle$ is expected to approach to the ideal state $\alpha|E_{0}\rangle|0\rangle+\beta|E_{1}\rangle|1\rangle$.
In practice, after the first twirling operation the physical state and the first ancilla bit will be in a state $c_{1}|\psi_{1}\rangle|0\rangle+d_{1}|
\varphi_{1}\rangle|1\rangle$, where $|\psi_{1}\rangle$ is a substitute of the true vacuum $|E_{0}\rangle$ and $|\varphi_{1}\rangle$ is an obstructive state
like the exited state $|E_{1}\rangle$.  The first approximate values of physical quantities are computed as the expectation values by $|\psi_{1}\rangle$.
We compute the second time parameter $\theta_{1}$ by $\theta_{1}E_{01}={\pi \over 2}$, where $E_{01}=\langle \psi_{1}|{\hat H}_{T}|\psi_{1}\rangle$.
Using $\theta_{1}$ and the second ancilla bit, we carry out the second twirling operation and obtain a state  $c_{2}|\psi_{2}\rangle|0\rangle|0\rangle+d_{2}|
\varphi_{2}\rangle||1\rangle\rangle$, where $||1\rangle\rangle$ means that at least one of the ancilla qubits is in the state $|1\rangle$.  We call the states that ancilla bits are 
in  $||1\rangle\rangle$ as excluded states.  While, we call the states that all of anncilla bits are in the state $|0 \rangle$ as active states.  Using the state $|\psi_{2}\rangle$, 
we update the values $E_{01}$ and $\theta_{1}$ to $E_{02}$ and $\theta_{2}$.    We can repeat this procedure as the number of the ancilla qubits.
\\  \\
Let us set $|\psi_{j}\rangle=\alpha_{j}|E_{0}\rangle+\beta_{j}|E_{1}\rangle$, $j=0,1,\cdots,m-1$, where $\alpha_{0}=\alpha, \beta_{0}=\beta$.
By the $j$-th twirling operation $|\psi_{j}\rangle$ and the $j$-th ancilla bit $|0\rangle$ transforms as
\begin{eqnarray}
|\psi_{j}\rangle|0\rangle \rightarrow ({1+ie^{-i\theta_{j}E_{0}} \over 2}\alpha_{j}|E_{0}\rangle+{1+ie^{-i\theta_{j}E_{1}} \over 2}\beta_{j}|E_{1}\rangle)|0\rangle  \nonumber \\
+ ({1-ie^{-i\theta_{j}E_{0}} \over 2}\alpha_{j}|E_{0}\rangle+{1-ie^{-i\theta_{j}E_{1}} \over 2}\beta_{j}|E_{1}\rangle)|1\rangle.
\end{eqnarray}
Since $\theta_{j}E_{0j}={\pi \over 2}$ and $E_{0j}$ is an approximate value of $E_{0}$, we have $\theta_{j}E_{0} \simeq {\pi \over 2}$.   Therefore,
it is expected that 
$|{(1+ie^{-i\theta_{j}E_{1}})\beta_{j} \over (1+ie^{-i\theta_{j}E_{0}})\alpha_{j}}|<|{\beta_{j} \over \alpha_{j}}|$ and the state corresponding to all of
the ancilla bits are in the state $|0\rangle$ approaches to the true vacuum $|E_{0}\rangle$.
\subsection{Simulation results }
In Table I we exhibit our simulation results for the case $J=1$.\\
\\
\begin{center}
\begin{tabular}{|c|c|c|c|c|c|c|} \hline
   &   0   &   1  & 2 & 3 & 4 & 5 \\ \hline
$10 \times 10^{6}$ &-0.71243 &-0.70739
 &-0.70702 & -0.70719& -0.70721&-0.70720 \\ \hline
$10 \times 10^{7}$   & -0.71234   &  -0.70703  & -0.70707 & -0.70714  & -0.70719& -0.70703 \\ \hline
\end{tabular}
\end{center}
{} \qquad 
\\ 
Table I \qquad Simulation results of the expectation value of $Z$ by IBM Qiskit qasm-simulator at the end of the adiabatic quantum computation($s=1$).  The numbers $0 \sim 5$ mean the number of times
of the twirling operation.    The number $10 \times 10^{6}(10 \times 10^{7})$ means that we have averaged over $10$ times for
each $10^{6}(10^{7})$ trials.  One twirling operation has been carried out by 100 steps.\\
\\
\\ 
The variance of measurement values of $Z$ for the state $|E_{0}\rangle$ is $\sigma^{2}={1 \over 2}$.   The statistical error of the expectation value of
$Z$ for $n=10^{6}$ trials is $\pm {\sigma \over \sqrt{n}}=\pm 0.0007$.    Therefore, roughly speaking, the expectation value of $Z$ for ten times average over $n=10^{6}$ trials
 will mainly be distributed over the range $-0.7073 \sim -0.7069$.   In this point our simulation results agree with this theoretical result.  For $n=10^{8}$ trials
$\pm {\sigma \over \sqrt{n}}=\pm 0.00007$ and the main range of the distribution will be $-0.70719 \sim -0.70704$.

Suppose after the second twirling operation we have the following ideal state that the ground state and the exited state are completely
separated
\begin{equation}
\alpha|E_{0}\rangle|0\rangle|0\rangle+\beta|E_{1}\rangle||1\rangle\rangle,
\end{equation}  
where the symbol $||1\rangle\rangle$ indicates that at least one $|1\rangle$ state is included  in the ancilla bits.
For this state we can obtain vacuum expectation values precisely from the first term.   Adding an ancilla bit in the state $|0\rangle$
to this state, we carry out the third twirling operation between the first bit and this ancilla bit.    After the third twirling operation, a quasi quantum state
in the classical computer that corresponds to the state $|E_{1}\rangle||1\rangle\rangle|0\rangle$ is decided to be $|E_{1}\rangle|0\rangle|0\rangle|0\rangle$
with small but non-negligible probability.   This will cause the increase of the number of active states and the expansion of discrepancy of vacuum expectation
values from the exact values.   Theoretically, once ancilla bits are excluded to the state $||1\rangle\rangle$, this state cannot revive  to the state
$|0\rangle|0\rangle{\cdots}|0\rangle$.   In the Qiskit qasm-simulator, measurements do not affect the quasi quantum state in the classical computer, 
and this situation does not hold.   It may be appropriate to decide the value as a candidate of the vacuum expectation value when the number of active states
is the smallest after some number of twirling operations.

We show in Table II our simulation results of the vacuum expectation value of $Z$ and the number of active states for a series of each round of the twirling
operation for $n=10^{8}$ trials. Except for the no twirling operation case the biggest deviation from the exact value $-{\sqrt{2} \over 2}$ appears for the
5 times twirling operation case, which is $1.71{\sigma \over \sqrt{n}}$.

\begin{center}
\begin{tabular}{|c|c|c|c|c|c|c|} \hline
   &   0   &   1  & 2 & 3 & 4 & 5 \\ \hline
$10^{8}$ &-0.71241 &-0.70704
 &-0.70706 & -0.70707& -0.70712&-0.70699 \\ \hline
active state   & 100000000  &  99998535  & 99998549 & 99998547  & 99998442& 99998525 \\ \hline
\end{tabular}
\end{center}
Table II \qquad An example of vacuum expectation value of $Z$ for $J=1$ and the number of active states for each round of the twirling
operation.   The numbers $0,1,\cdots,5$ indicate the number of times that we have acted the twirling operation.  We have used $n=10^{8}$ shots.\\

Theoretically the number of the active states should monotonically decrease with respect to the number of times of the twirling operation.
The simulation result shows this does not hold.   We are using Qiskit qasm-simulator on classical computers and measurements of states are supposed to be done in some sense classically.   
In quantum theory, the excluded states in $||1\rangle\rangle$ do not come back to be the active states,
as far as no operations have been done except for the specific qubits.
In qasm-simulator, however , this does not seem to hold.

Let us consider, for example, measuring the quntum state $|+\rangle={1 \over \sqrt{2}}|0\rangle+{1 \over \sqrt{2}}|1\rangle$ in the basis $\{|0\rangle,|1\rangle\}$.
In the actual quantum measurement we get the state $|1\rangle$ with the probability ${1 \over 2}$ and the state deterministically remains to be $|1\rangle$
by the following quantum measurements in the basis  $\{|0\rangle,|1\rangle\}$, if no operations are acted on the state.
In contrast with this, in the classically emulated quantum measurement in the basis  $\{|0\rangle,|1\rangle\}$ by the qasm-simulator, although once the 
$|+\rangle$ state is measured to be $|1\rangle$, the state may be measured to be $|0 \rangle$ in the following measurement.   This will be the reason
why the number of the active state does not monotonically diminish in our simulation.    In this situation, in our classically emulated quantum simulation
increasing the number of times of the twirling operation does not necessarily lead to the improvement of the vacuum expectation value of $Z$.

 \section{ (1+1)-dimensional Schwinger model on a one-dimensional spatial lattice}
The (1+1)-dimensional Schwinger model with the $\theta$-term is described in the natural unit of system $\hbar=c=1$ by the following Lagrangian density  
\begin{equation}
{\cal L}=-{1 \over 4}F_{\mu\nu}F^{\mu\nu}+{g\theta \over 4\pi} \epsilon_{\mu\nu}F^{\mu\nu} +i{\bar \psi}\gamma^{\mu}(\partial_{\mu}+igA_{\mu})\psi-m{\bar \psi}{\psi},
\end{equation}
where $\gamma^{0}=\sigma^{3}= \left(\begin{array}{cc}
                          0 & 1 \\
                         1  &  0 
\end{array} \right)$,  $\gamma^{1}=i\sigma^{2}= \left(\begin{array}{cc}
                          0 & 1 \\
                         -1  &  0 
\end{array} \right)$, $F_{\mu\nu}=\partial_{\mu}A_{\nu}-\partial_{\nu}A_{\mu}$.   We use the metric $(g_{\mu\nu})=\left(\begin{array}{cc}
                          1 & 0 \\
                          0 &  -1 
\end{array} \right)$ and we set the dielectric constant of vacuum as $\epsilon_{0}=1$.   We are interested in the simplest case $m=0,\theta=0$.  In the temporal gauge $A_{0}=0$, the corresponding Hamiltonian is
\begin{equation}
{\hat H}=\int dx(-i{\bar \psi}\gamma^{1}(\partial_{1}+igA_{1})\psi+{1 \over 2}\Pi^{2}),
\end{equation}
where $\Pi={\dot A^{1}}=-E^{1}$.  To investigate vacuum of this Hamiltonian by classically emurated quantum simulation, we formulate this Hamiltonian on a one-dimensional
spatial lattice with lattice spacing $a$.  To avoid the species doubling problem, we introduce Susskind fermions\cite{Susskind} with one-flavor.
For $x=na$($n$=even), we set $\psi(x)=(\psi_{n} \; 0)^{T}={1 \over \sqrt{a}}(\chi_{n} \; 0)^{T}$.  For even $n$, $\chi_{n}$ represents an annihilation operator of a particle and 
$\chi_{n}^{\dagger}$ represents a creation operator of the particle on the site $x=na$.   For $x=na$($n$=odd), we set $\psi(x)=(0\; \psi_{n})^{T}={1 \over \sqrt{a}}(0\; \chi_{n})^{T}$.  For odd $n$, $\chi_{n}^{\dagger}$ represents an annihilation operator of the antiparticle and 
$\chi_{n}$ represents a creation operator of the antiparticle on the site $x=na$.  These fermionic operators satisfy the following anti-commutation relations
\begin{equation}
\{\chi_{n}^{\dagger},\chi_{m}\}=\delta_{nm}, \qquad \{\chi_{n},\chi_{m}\}=0.
\end{equation}
We introduce the link variable $L_{n}$ by $L_{n}=-{1 \over g}\Pi(x=(n+{1 \over 2})a)$ that lives on the link connecting the sites $x=na$ and $x=(n+1)a$.
Thus we have lattice Hamiltonian, which leads to Eq.(10) in the limit $a \rightarrow 0$, as
\begin{equation}
{\hat H}=-iw\Sigma_{n=-\infty}^{\infty}(\chi_{n}^{\dagger}e^{iagA_{1}}\chi_{n+1}-\chi_{n+1}^{\dagger}e^{-iagA_{1}}\chi_{n})+G\Sigma_{n=-\infty}^{\infty}L_{n}^{2},
\end{equation}
where $w={1 \over 2a}$, $G={1 \over 2}g^{2}a$. Rewriting $e^{iagA_{1}}\chi_{n+1}$ as $\chi_{n+1}$, which does not affect the anti-commutation relations Eq.(11), we have
\begin{equation}
{\hat H}=-iw\Sigma_{n=-\infty}^{\infty}(\chi_{n}^{\dagger}\chi_{n+1}-\chi_{n+1}^{\dagger}\chi_{n})+G\Sigma_{n=-\infty}^{\infty}L_{n}^{2}.
\end{equation}

\section{Two-qubits system based on the (1+1)-dimensional Schwinger model}
We consider the simplest case of Eq.(13); let us only have two sites $x=0$ and $x=a$ on the spatial lattice.   We also adopt
the fixed boundary condition; the electric field is zero out of the region $0 \le x \le a$\cite{Honda1}.    The Gauss law will hold on physical states.  
The discretized version of the Gauss law $\partial_{1}E^{1}=\rho=g:\psi^{\dagger}\psi:$
at $x=0$ is
\begin{equation}
L_{0}-L_{-1}=\chi_{0}^{\dagger}\chi_{0}.
\end{equation}
From the fixed boundary condition $L_{-1}=0$ and we have $L_{0}=\chi_{0}^{\dagger}\chi_{0}$. Thus we have the following Hamiltonian on the two sites
\begin{equation}
{\hat H}=-iw(\chi_{0}^{\dagger}\chi_{1}-\chi_{1}^{\dagger}\chi_{0})+G(\chi_{0}^{\dagger}\chi_{0})^{2}.
\end{equation}
Using the Jordan-Wigner transformation\cite{Jordan}, we represent the fermionic variables $\chi_{0}$ and  $\chi_{1}$ by spin variables
\begin{equation}
\chi_{0}={1 \over 2}(X_{0}-iY_{0}), \qquad \chi_{1}={1 \over 2}(-iZ_{0})(X_{1}-iY_{1}).
\end{equation}
Thus we have the following simplest lattice Hamiltonian \cite{Honda1} up to a constant
\begin{equation}
 {\hat H}={1 \over 2}GZ_{0}+{1 \over 2}w(X_{0}X_{1}+Y_{0}Y_{1}).
\end{equation}
The first term in Eq.(17) is the electric field energy and the second term in Eq.(9) corresponds to the fermionic kinetic energy.
For simplicity we study the following Hamiltonian
\begin{equation}
 {\hat H}={1 \over 2}(X_{0}X_{1}+Y_{0}Y_{1})+JZ_{0}, 
\end{equation}
where $J={G \over 2w}={1 \over 2} g^{2}a^{2}$.
The matrix representation of Eq.(18) is
\begin{eqnarray}{\hat H}= \left(\begin{array}{cccc}
          J & 0 & 0 & 0 \\
          0     & J & 1 & 0 \\
          0    & 1 & -J & 0 \\
          0    &  0 & 0 & -J
\end{array} \right)
\end{eqnarray}
The smallest eigenvalue of Eq.(19) is $E_{0}=-\sqrt{1+J^{2}}$ and the corresponding eigenstate is $|E_{0}\rangle={1 \over \sqrt{2(J^{2}+J\sqrt{J^{2}+1}+1)}} {}^{t}(0,1,-\sqrt{J^2{}+1}-J, 0)$.    We are interested in the vacuum expectation value of ${\bar Z} \equiv {1 \over 2}(Z_{0}-Z_{1})${\cite{Honda1}}.   we see that
$\langle E_{0}|{\bar Z}|E_{0}\rangle=-{J^{2}+J\sqrt{J^{2}+1} \over J^{2}+J\sqrt{J^{2}+1}+1}$, which is $-{1 \over \sqrt{2}}$ for $J=1$ and is $-{2 \over \sqrt{5}}$ for $J=2$. 
We show in Table III our simulation results of the vacuum expectation value of ${\bar Z}$ and the number of active states for a series of each round of the twirling
operation for $n=10^{8}$ trials. 
For $J=1$, the theoretical value of the variance of ${\bar Z}$ is $\sigma^{2}={2+\sqrt{2} \over 4}$ when the ground state $|E_{0}\rangle$ is measured by the basis $\{|0\rangle,|1\rangle\}$.    The standard deviation of the average of ${\bar Z}$ is ${\sigma \over \sqrt{n}}=0.000092\cdots$.    Therefore, except the no twirling case 
the average of ${\bar Z}$ is within one standard deviation from the theoretical value $-{1 \over \sqrt{2}}$.\\
For $J=2$, the theoretical value of the variance of ${\bar Z}$ is $\sigma^{2}={1 \over 5}$ when the ground state $|E_{0}\rangle$ is measured by the basis $\{|0\rangle,|1\rangle\}$.    The standard deviation of the average of ${\bar Z}$ is ${\sigma \over \sqrt{n}}=0.000045\cdots$.    Therefore, except the no twirling case 
the average of ${\bar Z}$ is within two standard deviations from the theoretical value $-{2 \over \sqrt{5}}=-0.8994427\cdots$.
\\
\begin{center}
\begin{tabular}{|c|c|c|c|c|c|c|c|} \hline
   &   0   &   1  & 2 & 3 & 4 & 5 & 6\\ \hline
$10^{8}(J=1)$ &-0.71475 &-0.70716
 &-0.70701 & -0.70708& -0.70703&-0.70706 &-0.70710\\ \hline
active state   & $10^{8}$  &  99996822  & 99996734 & 99996858  & 99996733 & 99996846&99996864 \\ \hline \hline
$10^{8}(J=2)$ &-0.90014 &-0.89438
 &-0.89435 & -0.89440& -0.89442&-0.89441 &-0.89439\\ \hline
active state   & $10^{8}$  &  99995612  & 99995742 & 99995488  & 99995518 & 99995548 &99995709 \\ \hline
\end{tabular}
\end{center}
{} \quad \\
Table III\qquad An example of vacuum expectation value of ${\bar Z}$ and the number of active states for each round of the twirling
operation for J=1 and J=2.   The numbers $0,1,\cdots,6$ indicate the number of times that we have acted the twirling operation.  We have used $n=10^{8}$ shots.
The initial Hamiltonian is ${\hat H}_{0}= {1 \over 2}(Z_{0}-Z_{1})$ and we have started from the initial ground state $|1\rangle_{0}|0\rangle_{1}$.

\section{Summary and discussions}
We have proposed a procedure to extract vacuum expectation values from the approximate vacuum prepared by the adiabatic quantum computation.
We have developed the previous method presented by the present author\cite{Oshima} by concatenating the ancilla bits and the twirling operations.
By the concatenation it is expected that the approximate vacuum approaches the true vacuum more closely. 
We have carried out classically emulated quantum simulation by using IBM Qiskit for the one-qubit system and the two-qubits system that is
based on the (1+1)-dimensional Schwinger model.   We have seen that the vacuum expectation values we have obtained by the simulation
agree with the theoretical values up to the statistical errors.   Our simulation is not entirely quantum.    Therefore, the number of the active states
does not monotonically decrease as for number of the twirling operations in contrast with the case of entirely quantum simulation.
If we could use a noiseless quantum computer the number of the active states would monotonically decrease.    In this case before
computing vacuum expectation values we should obtain an approximate value of $E_{0}$, which can be done using repeatedly the same
Qiskit program.

We have simulated the one-qubit system, which has two energy eigenstate.   We also have simulated the two-qubits system based on the
(1+1)-dimensional Schwinger model.    Although this system has four energy eigenstates, since we have started from the charge neutral
state and the Hamiltonian conserves the number of the charge, the approximate vacuum is supposed to be a superposition of
two charge neutral states; the state no particles are exited and the state a pair of an electron and a positron is exited.
Therefore, concerning about the vacuum, this system is substantially a two quantum states system.    In our simulation results
we have obtained satisfactory vacuum expectation values by only one twirling operation.    This may be peculiar to a
two quantum state system.   We cannot have recognized benefit of concatenated ancilla qubits.    For an approximate
vacuum that contains plural excited states, our procedure may be effective.   To decide it by simulating more complex systems
is beyond our scope in this paper.

\newpage
Figure captions\\
\\
Fig.1   Vacuum expectation value of $Z$ for $J=1$ by IBM Qiskit qasm-simulator.  The parameter $0 \le s \le 1$ is
mapped to the time period $0 \le t \le 36$ and one time-step is ${1 \over 24}$.   We use $10^{6}$ shots.    The time
period $36 \le t \le 72$ exhibits the time evolution by the constant Hamiltonian ${\hat H}_{T}$.   We have used
the second order Suzuki-Trotter formula\cite{Trotter,Suzuki}.   We have started from the initial state $|0\rangle$.\\
\\
Fig.2(a)  One-qubit $j$-th twirling operation.   The operator $U_{j}$ is defined by $U_{j}=ie^{-i{\theta_{j}}{\hat H}_{T}}$, where $\theta_{j}$ 
satisfies $\theta_{j}E_{0j}={\pi \over 2}$ for $E_{0j}$ that is computed by $E_{0j}=\langle \psi_{j}|{\hat H}_{T}|\psi_{j}\rangle$.\\
\\
\\
Fig.2(b) One-qubit twirling operations with $m$ steps.\\
\\
\\
Fig.3 Two-qubit twirling operations with $3$ steps.\\
\newpage
{\quad } \includegraphics[width=8cm]{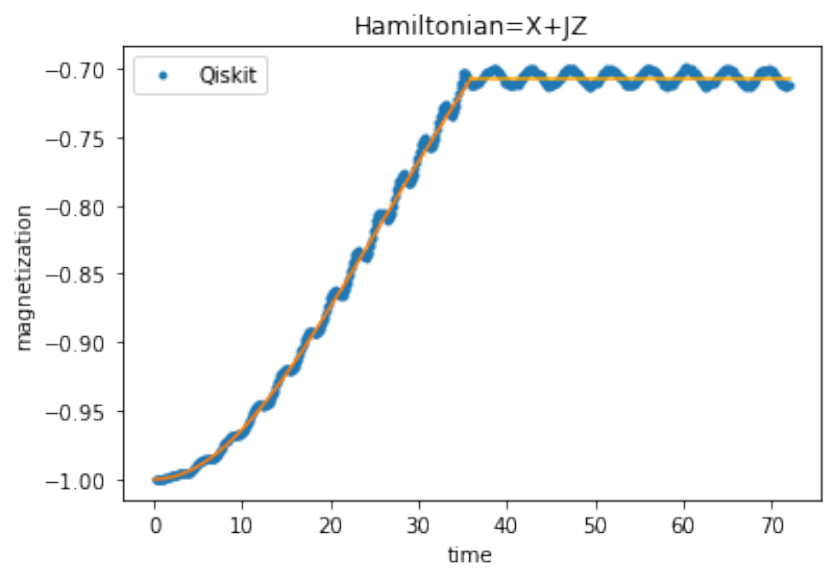}
\\
Fig.1\\
\\
\vspace{10pt}
{\quad }  \includegraphics[width=8cm]{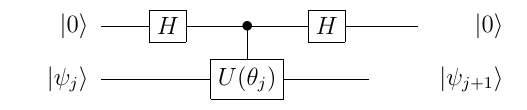}
\\
Fig.2(a)\\
\vspace{10pt}
\\ \\ \\
{\qquad}  \includegraphics[width=10cm]{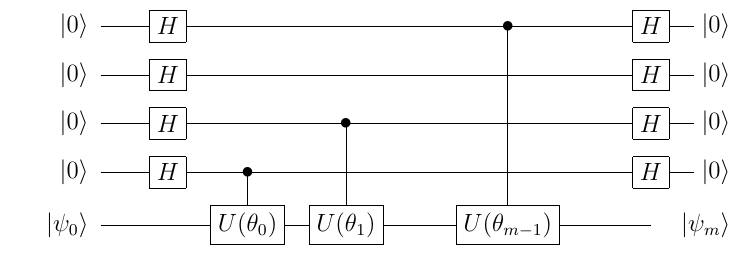}
\\
Fig.2(b)\\
\vspace{10pt}
\\ \\ \\
{\qquad}  \includegraphics[width=10cm]{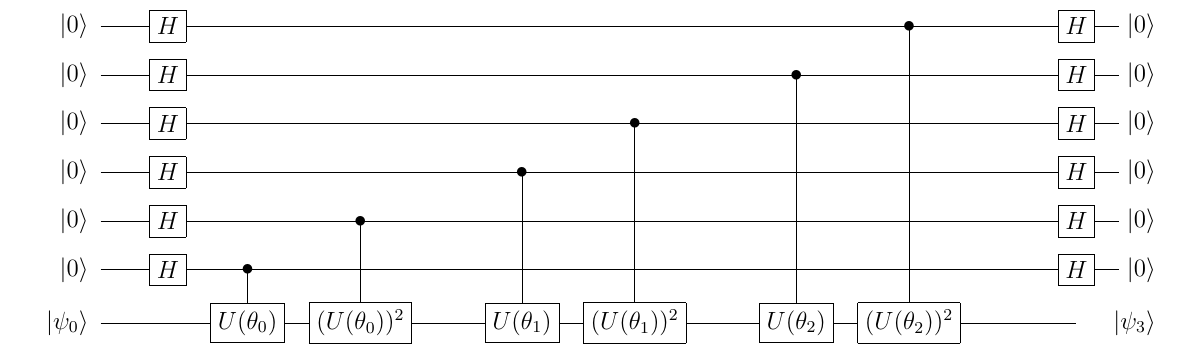}
\\
Fig.3\\
\vspace{10pt}
\\

\newpage
\begin{thebibliography}{99}

\bibitem{Born}
M. Born and V. A. Fock, Zeitschrift für Physik A. {\bf 51},165 (1928). \\
 T. Kato, Journal of the Physical Society of Japan. {\bf 5}, 4358(1950).

\bibitem{Nishimori}
T.Kadowaki and H.Nishimori, Phys.Rev.E{\bf 58}, 5355(1998).

\bibitem{Farhi1}
       E.Farhi, J.Goldstone, S.Gutmann, J.Lapan, A.Lundgren and D.Preda, Science{\bf 292}, 472(2001).

\bibitem{Martinez1}
E.A.Martinez, C.A. Muschik, P.Schindler, D.Nigg, A.Erhard, M.Heyl, P.Hauke, M.Dalmonte, T.Monz, P.Zoller and R.Blatt, {\bf 534}, 516 (2016).


\bibitem{Schwinger}
{ J.S.Schwinger, Phys.Rev.{\bf 125}, 397(1962). }

\bibitem{Honda1}
B.Chakraborty, M.Honda, T.Izubuti, Y.Kikuchi and A.Tomiya, Phys. Rev. D{\bf 105}, 094503 (2022) .

\bibitem{Honda2}
M.Honda.E.Itou, Y.Kikuchi and Y.Tanizaki, PTEP{\bf 2022}, 033B01(2022).

\bibitem{Okuda}
 M.Honda, E.Itou, Y.Kikuchi, L.Nagano and T.Okuda, Phys. Rev. D{\bf 105}, 014504 (2022).

\bibitem{Coleman}
{ S.R.Coleman, Annals Rev.{\bf 101}, 239(1976). }

\bibitem{Adam}
{ C.Adam, Phys.Lett.B{\bf 440}, 117(1998).}

\bibitem{Oshima}
{K.Oshima, IET Quant. Comm., 1(2022)}

\bibitem{Trotter}
H.H.Trotter, Proc. Amer. Math. Soc. {\bf 10} ,545(1959).

\bibitem{Suzuki}
M.Suzuki, Comm. Math. Phys. {\bf 51}, 183 (1976). 


\bibitem{Susskind}
{ L.Susskind, Phys.Rev.{\bf 16}, 3031(1977). }

\bibitem{Jordan}
P.Jordan and E.Wigner, Z.Phys. A{\bf 47}, 631(1928).







\end{thebibliography}
\end{document}